# The core modes of all solid photonic band gap fibers as bound states in the continuum

Andrey D. Pryamikov [1,]*

[1] Fiber Optics Research Center of Russian Academy of Sciences, 119333, 38 Vavilov str., Moscow, Russia; E-Mails: pryamikov@fo.gpi.ru (F.L.)

* Author to whom correspondence should be addressed; E-Mail: pryamikov@fo.gpi.ru (F.L.); Tel.: +7-499-503-8193; Fax: +7-499-135-8139.

**Abstract:** In this paper we propose a new hypothesis of out of plane band gap formation in all solid photonic band gap (PBG) fibers. At first, one considers the process of the core modes formation as a result of excitation of continuous mode spectrum of single dielectric rod of the cladding. It is shown that the core modes can be considered as transverse resonance of a set of the radiation modes decoupled from the continuum provided that the phase matching condition for the propagation constants are fulfilled. An analogy between the core mode formation in all solid PBG fibers and formation of bound states in the continuum in photonics are drawn. Based on these conclusions, the light guiding in all solid PBG fiber with disordered cladding structure and quasi 3D photonic band gap formation in all solid PBG fibers with a small index (1%) can be explained.

## 1. Introduction

All solid photonic band gap (PBG) fibers have attracted considerable attention recently due to their useful and interesting properties [1]. In the case of all solid PBG fibers light propagates in a low index defect in contrast to holey fibers, which guide light through the modified total internal reflection [2]. Solid core PBG fibers consist of an array if high index cylinders in a low index background and a low index defect – core in which light propagates. These fibers have transmission spectra with several low loss windows. Light is confined to the core by a photonic bandgap of the cladding. In the limit of infinite, periodic microstructure of the cladding the modal properties of solid core PBG fibers can be modeled by Bloch's theorem and plane wave expansion [1, 3]. The second approach to modeling the modal properties of all solid PBG fibers is connected with the resonant properties of an isolated high index rod of the microstructured cladding [4]. The authors of [4] showed the confinement losses in all solid PBG fibers depend strongly on wavelength. At the same time, the positions of loss minima and maxima depend on the scattering properties of individual cylinders and weakly on their position in the cladding and their number. It means that it is not necessary to have periodically arranged rods in the cladding of all solid PBG fibers for the light localization in the defect – core. On the other hand, it is known that very large refractive index contrast is needed to create a bandgap. For example, full 2D PBG, when the light propagates perpendicular to an array of rods, can appear at refractive index contrast no less than 2.66 [5] but nonzero component parallel to 2D PBG structure allow to obtain bandgap at much smaller refractive index contrasts [6]. In work [7] the authors experimentally demonstrate the existence of photonic bandgaps for arbitrarily small refractive index step, in particular, for all solid PBG fiber with a relative index step of 1%. The authors of [7] pointed out that this fact contradicts the conclusions made in early works on photonic bandgaps, large refractive index contrast is not necessary for photonic crystal guiding in all solid PBG fibers.

In this paper we shall try to explain the results obtained in [4] and [7] based on a new approach connected with bound states in the continuum in photonics [8].

## 2. Results and Discussion

In this section we consider two approaches to an explanation of waveguide mechanism in all solid PBG fibers. In the first subsection we will try to describe the core modes formation in all solid PBG fibers based on continuous mode spectrum of circular dielectric rods of the cladding. In the second subsection we will try to justify such approach by considering a process of formation of bound state in the continuum in photonics.

*2.1. The continuous mode spectrum of a circular dielectric rod and the core mode formation in all solid PBG fiber*

The first question is that why it is not necessary to have an accurate hexagonal arrangement of the rods in the cladding to form the core modes of all solid PBG fiber [4]? The second question is that why it is possible to open up quasi 3D bandgaps in the case of SC PBG with such a small refractive index step as in [7]?

Up to this time all resonant effects occurred under waveguide propagation in all solid PBG fibers have been considered under the core modes interaction with high order leaky modes of the individual rods of the cladding at the cutoffs of these high order modes of the rods [9]. As a result of this interaction, the special features in the spectral dependencies of the real parts of the propagation constants of the core modes can be observed [10]. In work [11] Fano type resonances were observed under the core mode interactions with leaky modes of the rods in the case of small values of the ratio $d/\Lambda$, where $d$ is a diameter of the cladding rod and $\Lambda$ is a pitch. The waveguide characteristics of the core modes of all solid PBG fibers and the waveguide characteristics of the mode content of the single dielectric rod of the cladding have been considered separately from each other. The coupling between them occurs just near the cutoff wavelengths. In this connection, the question arises: in what way the core modes of all solid PBG fibers are formed? Is there any connection between the core mode and the mode content of the single dielectric rod of the photonic crystal cladding away from the resonant wavelengths (the cutoff wavelengths)?

In the case of Bragg fibers and other microstructured fibers having uniaxial rotational symmetry and 1D translational symmetry the answer is simple: the electromagnetic fields of each core mode of *m*th order (*m* = 0, 1 …) have the general form in cylindrical coordinate:

$$\vec{E}(r,\varphi,z) = A(\beta_m)\vec{F}_m^e(r,\varphi)e^{-i\beta_m z} \tag{1}$$

$$\vec{H}(r,\varphi,z) = B(\beta_m)\vec{F}_m^h(r,\varphi)e^{-i\beta_m z},$$

where $\vec{F}_m^e$ and $\vec{F}_m^h$ are the transverse dependencies of the electric and magnetic fields. The propagation constant $\beta_m$ is determined from the corresponding boundary conditions for *z* and $\varphi$ - components of electric and magnetic fields. Then, the total electric field of the propagating modes in Bragg fiber can be written as:

$$\vec{E}^{total} = \sum_{m=1}^{\infty} A(\beta_m)\vec{F}_m^e(r,\varphi)e^{-i\beta_m z}. \tag{2}$$

In such a way, each core mode of Bragg fiber is formed by only one cylindrical harmonic and is independent on cylindrical harmonics of different orders which form the other core modes. These core modes can be considered as transverse resonances of electromagnetic fields of individual cylindrical harmonics in the core. For example, *z* – component of the electric field of *m*th order core mode can be expressed as:

$$E_z = \left\{A_m J_m(k_{in}^m r) + B_m H_m^{(2)}(k_{in}^m r)\right\} \begin{Bmatrix} \cos(m\varphi) \\ \sin(m\varphi) \end{Bmatrix} e^{-i\beta_m z}, \tag{3}$$

where $k_{in}^m = \sqrt{k_i^2 - \beta_m^2}$ is a transverse wavevector component and $k_i = 2\pi n_i / \lambda$ is a wavevector in *i*th layer. Such propagating core mode transfers a momentum of $\beta_m$ in *z* – direction and azimuthal dependencies (3) are represented in the form of 'standing waves' (no momentum of ligth is transferred in $\varphi$ - direction). It is also worth mentioning that the eigenvalue $\omega^2(k_{in}^m)$ depends only on the absolute value of $k_{in}^m$ and not on its direction as in all other systems with continuous rotational symmetry [12]. In this case, the resonances which can be excited in the cladding layers are the radial resonances when a standing wave is formed in the high index layers and they can be considered as Fabry – Perot resonators. Such resonances are described by the ARROW model [13].

In the case of all solid PBG fibers the core mode formation is much more complicated process when the fields of each core mode are represented by infinite series of cylindrical harmonics. It is known that one of the most effective numerical methods for calculation of the core modes characteristics is multipole method [14]. The method uses the circularity of the cladding inclusions and can treat precisely the systems containing large number of inclusions. Let us consider an extreme case of all solid PBG fiber with small value of $d/\Lambda$ [11, 15]. In this case, it can be stated that the coupling between the rods of the cladding is weak [11]. Each *m*th order core mode of all solid PBG is also characterized by its propagation constant $\beta_m^{core}$ and by the corresponding transverse dependencies of the fields [14]:

$$\vec{E}(r,\varphi,z) = A(\beta_m^{core})\vec{F}_{core,m}^e(r,\varphi,\beta_m^{core})e^{-i\beta_m^{core} z} \tag{4}$$

$$\vec{H}(r,\varphi,z) = B(\beta_m^{core})\vec{F}_{core}^h(r,\varphi,\beta_m^{core})e^{-i\beta_m^{core}z}.$$

According to work [14] all electromagnetic fields in the silica matrix in the vicinity of *l*th individual cylindrical inclusion are expressed in Fourier – Bessel series in the local coordinate system. For example, *z* – component of electric field is represented as:

$$E_z = \sum_{n=-\infty}^{+\infty}\{A_n^l J_n(kr) + B_n^l H_n^{(2)}(kr)\}e^{-in\varphi}e^{-i\beta z} \quad (5)$$

where $k = \sqrt{k_m^2 - \beta^2}$ is a transverse wavevector and $k_m = 2\pi n_m/\lambda$ is a wavector in the silica matrix. In particular, the representation (5) should be true for *z* – component of electric field of *m*th order core mode if $\beta = \beta_m^{core}$ and the transverse wavevector $k_{in} = \sqrt{k_m^2 - \beta_m^{core2}}$. On other hand, it is stated [14] that expansion (5) is equivalent to the expansion which was proposed in [16] where all electromagnetic fields originated from the neighboring cylinders are summarized in the vicinity of *l*th inclusion:

$$E_z = \left\{\sum_{l=1}^{N}\sum_{n=-\infty}^{+\infty} B_n^l H_n^{(2)}(kr)e^{-in\varphi} + \sum_{n=-\infty}^{+\infty} A_n^0 J_n(kr)e^{-in\varphi}\right\}e^{-i\beta z} \quad (6)$$

where the first series are connected with the outgoing waves from all the rods – inclusions *l* = 1, …, *N* and the final one is connected with regular field originated at the jacket boundary.

At the same time, it is known that each single dielectric rod with higher refractive index than the matrix has an infinite set of continuous modes (radiation modes) [17]. Such consideration of the radiation modes is equivalent to the problem of plane wave scattering from the dielectric rod. For example, ITM (incident transverse magnetic) continuous (radiation) modes occur when the longitudinal component of the incident plane wave $H_z = 0$ and $E_z \neq 0$. In such a way, *z* –component of electric field of ITM radiation mode of *n*th order can be written as [17]:

$$E_z = \{A_n J_n(k_m r) + B_n H_n^{(2)}(k_m r)\}e^{-in\varphi}e^{-i\beta z}, \quad r > a \quad (7a)$$

$$E_z = C_n J_n(k_c r)e^{-in\varphi}e^{-i\beta z}, \quad r < a \quad (7b)$$

where *a* is a radius of the rod, the transverse wavevectors are $k_c = \sqrt{(2\pi n_c/\lambda)^2 - \beta^2}$ and $k_m = \sqrt{(2\pi n_m/\lambda)^2 - \beta^2}$, where $n_c$ and $n_m$ are the refractive indices of the rod and the matrix. In this approach $J_l$ is associated with the plane wave and $H_n^{(2)}$ term is associated with the resulting scattered field. It is seen that the individual terms in the expansion of the local fields in the case of multipole method (5) has the same construction. In the general case, the electric and magnetic fields of radiation mode of *n*th order can be constructed intuitively as [18]:

$$\vec{e}_n = \vec{e}_n^f + \vec{e}_n^s \quad (8)$$

$$\vec{h}_n = \vec{h}_n^f + \vec{h}_n^s,$$

where the indices *f* and *s* denote the fields of 'free space' and the scattered fields.

There is also another classical approach to determining the radiation modes of single dielectric rod [17]. In this case, the radiation modes are considered as the limit of fields in the dielectric rod with some finite radius surrounded by a cylindrical metal wall of larger radius $\rho$. When $\rho$ tends to infinity the complete set of normalized modes of the dielectric rod in free space is found. Such approach can be considered as very similar to the Winjgaard expansion (6) when *J* terms are originated at the jacket boundary (the metallic wall). Based on the facts mentioned above, one can reasonably suggest that the fields of the core mode of all solid PBG fibers are formed by an interference of the excited radiation modes of the individual cladding rods. Further, we shall try to justify our suggestion.

In the general case, the total fields of the propagating radiation modes of single dielectric rod can be represented as [18]:

$$\vec{E}^{rad} = \sum_{n=-\infty}^{+\infty} \int_0^{k_m} A_n(\beta) \vec{F}_n^e(r,\varphi,\beta) e^{-i\beta z} d\beta \qquad (9)$$

$$\vec{H}^{rad} = \sum_{n=-\infty}^{+\infty} \int_0^{k_m} B_n(\beta) \vec{F}_n^h(r,\varphi,\beta) e^{-i\beta z} d\beta ,$$

where $\vec{F}_n^e$ and $\vec{F}_n^h$ are the transverse structures of electric and magnetic fields as in the case of the core mode of Bragg fiber (1), (2). Besides, the fields of the radiation modes for the determined value of $\beta$ can be considered as Fourier expansion of the plane wave family which propagates at an angle $\theta$ with respect to the rod axis *z* [18]. In this case, the propagation constant can be represented as $\beta = k_m \cos\theta$, where $0 < \theta < \pi/2$.

The situation changes if there is a system of the dielectric rods which are arranged in some order in silica matrix forming a defect - core as in the case of all solid PBG fiber. It is known that the real part of the propagation constants of the core modes of all solid PBG fibers is conserved under the mode propagation in the core at some wavelength in the transmission zone. In other words, there is no any momentum transfer from the photonic crystal cladding to the core mode in *z* – direction under such propagation of the core mode. In such a way, it is possible to say that each mode propagates in the core with a propagation constant of $\beta^{core} = k_m \cos\theta_z$, where $\theta_z$ is an angle of incidence of the core mode. It is known from the calculations and experiments [11, 15] that the value of $\beta^{core}$ is very close to $k_m$ (silica line) for all solid PBG fibers with small value of $d/\Lambda$ and small step index. It means that the core mode is incident on the cladding rods at grazing angles $\theta_z$. Due to this fact, it is possible to state that there is a very narrow interval of values of $\theta_z \ll 1$ in which the waveguide regime is possible in considered spectral range. Moreover, if one assumes that the core modes are formed by the interference of the radiation modes it is necessary to fulfill the phase matching condition in *z* – direction between the propagation constants of the core modes and the propagation constants of the radiation modes $\beta^{core} = \beta$. So as, the propagation constants of the radiation modes form a continuum it is always possible to achieve such phase matching condition. Therefore, there must be some finite interval of the propagation constants of the radiation modes $\beta$ in which it is possible to achieve the phase matching condition with the core modes of all solid PBG fibers and the low integration limit in (9) must be determined by some value of $\beta^{critical}$ which is also close to the silica line.

In such a way, the radiation modes which can form the core modes of all solid PBG fiber must be excited in very narrow interval of their propagation constants $\beta$ which must correspond to the geometry of the fiber core and to the values of $\beta_m^{core}$ of the excited core modes. It corresponds to the concept of the source at infinity $\theta \to 0$ when individual modes of the dielectric rods are excited selectively [17]. Due to this fact, only little radiation modes energy is confined to the rod surface. In this case, the fields of the radiation modes of the single rod are transformed into the following manner:

$$\vec{E}^{rad} = \sum_{n=-\infty}^{+\infty} \int_{\beta^{critical}}^{k_m} A_n(\beta) \vec{F}_n^e(r,\varphi,\beta) e^{-i\beta z} d\beta \qquad (10)$$

$$\vec{H}^{rad} = \sum_{n=-\infty}^{+\infty} \int_{\beta^{critical}}^{k_m} B_n(\beta) \vec{F}_n^h(r,\varphi,\beta) e^{-i\beta z} d\beta .$$

Now, applying the phase matching condition described above to the expressions (10) it is possible to obtain a system of transverse resonances for the sum of the fields of the radiation modes in the defect – core of all solid PBG fiber. The occurrence of such transverse resonances means that some radiation modes with propagation constants from the spectral interval given in (10) satisfying the phase matching condition become discrete. In other words, they are decoupled from the continuum. The number of the discrete modes $M$ (the transverse resonances) in a given spectral range is determined by the geometry parameters of all solid PBG fiber. In such a way, if some radiation modes described by expression (10) satisfy the transverse resonance conditions in the defect - core of all solid PBG fiber the integrals in (10) are transformed into the discrete sequences for them:

$$\vec{E}^{rad}_{resonance} = \sum_{m=1}^{M} \sum_{n=-\infty}^{+\infty} A_n(\beta_m^{core}) \vec{F}_n^e(r,\varphi,\beta_m^{core}) e^{-i\beta_m^{core} z} \qquad (11)$$

$$\vec{H}^{rad}_{resonance} = \sum_{m=1}^{M} \sum_{n=-\infty}^{+\infty} B_n(\beta_m^{core}) \vec{F}_n^h(r,\varphi,\beta_m^{core}) e^{-i\beta_m^{core} z} ,$$

where $\beta_m^{core}$ is the propagation constant of $m$th order core mode. If one uses the general expression for the fields of $m$th order core mode (4) one can obtain the following equalities in the vicinity of $l$th cladding rod:

$$A(\beta_m) \vec{F}_m^e(r,\varphi,\beta_m^{core}) = \sum_{n=-\infty}^{+\infty} A_n(\beta_m^{core}) \vec{F}_n^e(r,\varphi,\beta_m^{core}) \qquad (12)$$

$$B(\beta_m) \vec{F}_m^h(r,\varphi,\beta_m^{core}) = \sum_{n=-\infty}^{+\infty} B_n(\beta_m^{core}) \vec{F}_n^h(r,\varphi,\beta_m^{core}) .$$

It is seen that the transverse dependencies of the fields of *m*th order core mode in the vicinity of *l*th cladding rod can be expressed through the sum of the resonant radiation modes. On the other hand, applying the general expression (8) to (11) it is possible to obtain the fields of 'free space' and the scattered fields separately for each *n*th order resonant radiation mode. In this case, the inner sum in (11), for example, for $z$ – component of electric field at a given value of *m* is equivalent to expression (5) for $z$ – component of electric field of *m*th order core mode in the case of multipole method.

In our opinion, a representation of the core mode formation as an interference of the radiation modes is the main difference between the waveguide mechanisms in the systems with discrete rotational symmetry (all solid PBG fibers, hollow core fibers with negative curvature of the core boundary) and with continuous rotational symmetry (Bragg fibers). In the first case, the core modes are formed by a superposition of several resonant radiation modes with determined value of $\beta_m^{core}$ which are selectively excited by the source at $\theta \to 0$. In the second case, the core modes are discrete initially and each core mode is formed just by one resonance cylindrical harmonic. In such a way, the core modes of the systems with continuous rotational symmetry are discrete leaky mode which cannot be formed by the modes from the continuous spectrum.

Note, that in all our reasoning we haven't used an approach connected with Bloch modes of the photonic crystal cladding. Now, it is possible to say why the localization of light in the core of all solid PBG fiber is possible even in the case of random distribution of the rods in the cladding [4]. Some number of the radiation modes with propagation constants from the interval $\beta^{critical} < \beta < k_m$ can always be excited. The main demand for the core mode formation is that the phase matching condition $\beta = \beta_m^{core}$ and condition for the transverse resonance in the defect - core must be fulfilled. In the same way it is possible to explain the formation of full photonic band gap in the case of all solid PBG fibers with a small index step [7].

In this case, the question has to be answered is when the photonic crystal cladding begin to play decisive role in the bandgap formation and what is the criteria of applicability of photonic crystal description in the case of all solid PBG fibers described above? We shall try to answer this question in the next subsection. Moreover, we shall try to justify the conclusions made above in the next subsection using another approach.

In the end of this subsection, it is worth mentioning that Snyder in his book [18] said that there is no physical sense in the individual radiation mode and they are some abstraction which is very useful for describing the radiation processes. Indeed, the number of such modes excited under oblique incidence of light on the rod is limited and they can manifest themselves when they are decoupled from the continuum under satisfying the phase matching condition and the transverse resonance condition in the defect - core of all solid PBG fibers (11).

*2.2. The bound states in photonics and the core mode formation in all solid PBG fibers*

The resonance phenomenon connected with decoupling of the radiation modes from the continuum under the core modes formation in all solid PBG fibers is very similar to the phenomena considered in work [8, 19]. In work [8] two simple photonic systems were considered. One of them was two arrays of identical dielectric gratings which was invariant in one direction, for example, *y* and had a period *D* in *x* – direction. The primitive reciprocal lattice vector was $G = 2\pi/D$. In the symmetric case of two equivalent diffraction gratings the symmetric and antisymmetric quasistationary states (trapped modes) are formed in the structure. They give rise to two distinct dispersion curves around the dispersion curve of the single grating structure. One of these curves has extremely narrow resonance. Such large amplification of the resonance field corresponds to the long lived quasistationary state. The quasistationary states decay into the zero diffraction channel only if the next condition (the first diffraction threshold) $k + k_x < G$ is fulfilled [8], where $k$ is the total wavevector and $k_x$ is its component along the *x* axis. It means that there are no any changes in $k_x$ and the periodic structure doesn't transfer the momentum to the light in this direction. When the value of $k$ allows to change $k_x$ by *lG*, where *l* is an integer (diffraction order), the more diffraction channels are open and the resonance width increases. It is pointed out in [8] that for a large distance *h* between the two arrays of gratings the interaction through evanescent fields in closed diffraction channels ($|lG - k_x| > k$) can be neglected. The same situation occurs in the case of the double array of identical dielectric cylinders [8]. The resonance located below the first diffraction threshold has very small width and the long lived trapped mode is connected with strong field enhancement. Then, the authors of [8] pointed out that the mode stabilization occur when the resonance frequency of the system consisting of two arrays of diffraction gratings matches closely the frequency of the standing wave in the structure. As a conclusion, it was said that such scattering resonance with vanishing width represents a bound state in the radiation continuum, where the electromagnetic field is trapped by the structure for very long time.

On the hand, in work [19] the structure consisting of a periodic array of thin dielectric cylinders with dielectric constants higher than that of surrounding media can be effectively used for second harmonic generation. It was shown that the most effective second harmonic generation occurs when both fundamental and second harmonic fields are in resonance with the trapped modes of the structure, provided that the phase matching condition between the fundamental and second harmonic waves is fulfilled. It was also pointed out that as in the case of the system consisting of two diffraction arrays [8] the reflection of incident light is low except for the spectral lines lying just below the opening of the new diffraction channels $k \pm k_x = lG$. In this case, the incident radiation is in resonance with the trapped modes of the structure. As a conclusion, it was said that to achieve the most efficient second harmonic generation it was necessary that both fundamental and second harmonic wave were in resonance with some trapped mode, in other words, to obtain the transverse resonance provided that the phase matching condition is fulfilled.

Let us apply the results mentioned above [8, 19] and results obtained in the previous subsection to an explanation of waveguide mechanisms in all solid PBG fibers. According to our assumption in the previous subsection, the integrals (9) describing the fields of radiation modes of the single dielectric rods are transformed into the sets (11) under introducing a defect – core in 2D array consisting of such rods. It is clear that the propagation constants of the resonant radiation modes correspond to the transverse resonances with the transverse wavevector $k_{in} = \sqrt{k_m^2 - \beta_m^{core2}}$, where $k_m = 2\pi n_m/\lambda$ and $n_m$ is a refractive index of the matrix. In this case, both conditions described in [19] are fulfilled and the core mode formation in all solid PBG fibers under interference of the resonance radiation modes is most effective.

As it was mentioned earlier, the real part of the propagation constant of the core modes $\beta_m^{core}$ is conserved under the core mode propagation. On other hand, each mode has a transverse component of the wavevector which can be affected by 2D photonic crystal cladding. For example, all solid PBG fiber with hexagonal lattice and with geometry parameters given in [4] has the primitive reciprocal lattice vector $\vec{G} = \frac{2\pi}{\Lambda}\left(\vec{x} \pm \frac{\sqrt{3}}{3}\vec{y}\right)$, where $G_x$ = 1.114e6 m$^{-1}$ and $G_y$ = 6.432e5 m$^{-1}$. The transverse wavevector of the fundamental mode at λ = 0.792 μm is $k_{in}$ = 5.45e5 m$^{-1}$. It is seen that the fundamental core mode is certainly below the first diffraction threshold in the case of $x$ – component of the reciprocal lattice vector $G_x - k_x > k_{in}$ and $k_x$ cannot be changed by interaction with 2D photonic crystal cladding. Nevertheless, the core mode is well localized [4] and it means that the core modes formation occurs due to the resonant radiation mode interaction which comes to occurrence of bound state in the continuum. At the same time, in the case of $y$ – component of the reciprocal lattice vector the first diffraction threshold $G_y = k_{in} + k_y$ can be overcome, in principle, and $k_y$ can be changed by 2D photonic crystal cladding. In this case the resonance width is larger as both zero and first decay channel are open [8]. If the value of $k_y$ is such that the inequality $G_y - k_y > k_{in}$ is true the core mode of all solid PBG fibers is formed purely as a bound state in the continuum.

In such a way, it is possible to state that the core mode formation in all solid PBG fibers with a small refractive index step for which the conditions described above are satisfied is not connected with constructive Bragg interference which occurs due to light scattering from 2D photonic crystal cladding. In this case, the core mode formation is a result of interference of the radiation modes which are decoupled from continuum as it was described in previous subsection. The core mode stabilization occurs when the propagation constants of the radiation modes of the single cylinder satisfy the phase matching condition and the transverse resonance condition when quasi standing wave occurs in the defect - core for the sum of the fields of the radiation modes (11). Due to this fact it is possible to explain why the all solid PBG fibers [4] with disordered structure of the cladding can transmit the light in the core. All diffraction channels of the photonic crystal cladding are closed except of zero order diffraction channel and the core mode occur due to interaction between resonant radiation modes of the individual rods of the cladding when the standing wave (transverse resonance) is formed in the defect - core. Due to that there is no any dependence of the band locations on the photonic crystal structure of the cladding [4].

2D photonic crystal cladding doesn't transfer any momentum to the modes in transverse plane if the core modes are formed below the first diffraction threshold. It means that all resonant features occurred in the propagation mode spectrum must be connected with resonance properties of single dielectric rod of the cladding. In their turn, these resonances are just connected with cutoff wavelengths of the single rod. In this case the localization of the core modes is possible even for a small refractive index step.

## 3. Conclusions

In our opinion, the formation of out of plane band gap and low loss guidance in all solid PBG fibers with a small refractive index step are possible due to an excitation of the radiation modes of the individual rods of the cladding. If their propagation constants satisfy the phase matching condition in the propagation direction and the transverse resonance condition for the defect – core modes of all solid PBG fibers they decoupled from the continuum. They interfere between each other and form the core modes of the fiber. Due to the same reason the localization of the defect – core modes is possible in the case of random arrangement of the rods in the cladding. These facts can be confirmed by comparing the values of reciprocal lattice vectors of the photonic crystal cladding with the value of the transverse component of the wavevectors. Such approach to an explanation of light localization in the defect – core of all solid PBG fibers can be useful for understanding low loss guidance and the core mode formation in hollow core fibers with negative curvature of the core boundary.

**Conflict of Interest**

"The authors declare no conflict of interest".